# MPPSOCGEN: A FRAMEWORK FOR AUTOMATIC GENERATION OF MPPSOC ARCHITECTURE

Emna Kallel, Yassine Aoudni, Mouna Baklouti and Mohamed Abid

Electrical department, Computer Embedded System Laboratory,
ENIS School, Sfax, Tunisia

***ABSTRACT***

*Automatic code generation is a standard method in software engineering since it improves the code consistency and reduces the overall development time. In this context, this paper presents a design flow for automatic VHDL code generation of mppSoC (massively parallel processing System-on-Chip) configuration. Indeed, depending on the application requirements, a framework of Netbeans Platform Software Tool named MppSoCGEN was developed in order to accelerate the design process of complex mppSoC. Starting from an architecture parameters design, VHDL code will be automatically generated using parsing method. Configuration rules are proposed to have a correct and valid VHDL syntax configuration. Finally, an automatic generation of Processor Elements and network topologies models of mppSoC architecture will be done for Stratix II device family. Our framework improves its flexibility on Netbeans 5.5 version and centrino duo Core 2GHz with 22 Kbytes and 3 seconds average runtime. Experimental results for reduction algorithm validate our MppSoCGEN design flow and demonstrate the efficiency of generated architectures.*

***KEYWORD***

***MppSoC, Automatic code generation; mppSoC configuration;parsing ; MppSoCGEN;***

## 1. INTRODUCTION

Parallel machines are most often used in many modern applications that need regular parallel algorithms and high computing resources, such as image processing and signal processing. Massively parallel architectures, in particular Single Instruction Multiple Data (SIMD) systems, have shown to be powerful executers for data-intensive applications [1]. Indeed, the SIMD architectures are widely recognized as being well-suited for media-centric applications since they can efficiently exploit massive data parallelism available with minimal energy [2].

However, this class of processors has been less used in the nineties because of its dramatically-high fabrication cost [30]. Currently, the recent great progress of silicon integration technology on the one hand, and the wide usage of reusable Intellectual Property (IP) cores on the other hand, are more adopted to rise these defies and to reduce the time to market. Several SIMD architectures have been proposed to exploit data intensive parallelism [4], [15]. Although their implementation is still a problem, they pose a serious challenge for system developers. To accelerate and facilitate the development of these large and complex Systems-on-Chip (SoCs), new design methodologies need to be adopted.

Automation is the key to reduce design and verification time [13] as it allows system developers to traverse the design space in a much shorter time. Some automatic code generation tools for embedded software exist already. In fact, several commercial behavioral or high-level synthesis tools are available, e.g., Cynthesizer by Forte Design Systems [25], CatapultC by Mentor Graphics [26], and NEC's CyberWorkBench [27]. Moreover, automatic code generation academics tools for embedded processors are presented in [3, 9, 14]. But, to the best of our knowledge, there is no full code generation tool for massively parallel architectures.

In this paper, we address the above-mentioned problems and present a design flow based on automatic generation techniques to automatically generate complete VHDL code of mppSoC configuration. The basic idea is to make mppSoC configuration decisions in early design phases at high abstraction levels to help the designer select the best design alternatives. We prove our design flow effectiveness by implementing MppSoC generator tool (MppSoCGEN). MppSoCGEN is a java Netbeans based framework, which generates SIMD mppSoC configuration from a Graphical User Interface, in which the user can specify the number and type of PEs and their communication infrastructure. The MppSoCGEN design flow is used to generate several designs that have been tested on FPGA Stratix 2S180 device [24].

As a testament to the maturity of the MppSoCGEN design flow, we consider the reduction algorithm where we explore different parallel system configurations and decide on the best one to run the application. Experimental results show that the proposed framework considerably reduces design costs comparing to other hardware generator tool. Moreover, it facilitates mppSoC configuration and regenerating the implementation without relying on costly re-implementation cycles. Using the framework, we can create SIMD implementations that are fast enough to meet demanding processing requirements.

The paper is organized as follows. Section 2 discusses some significant works related to automatic generation approaches to generate on-chip multi-processor or massively parallel systems. Section 3 describes the mppSoC system and focuses on the proposed IP based assembling methodology to implement one mppSoC configuration on FPGA. It also introduces the developed MppSoCGEN framework. In Section 4, our design flow to the automatic mppSoC configuration is described. A case study illustrating and validating the framework is described in Section 5. The FPGA platform is chosen as a target platform to test and implement various parallel SoC configurations. Finally, section 6 draws main conclusions and proposes future research directions.

## 2. RELATED WORK

Multi-Processor Systems-on-Chip (MPSoCs) are increasingly popular in embedded systems. Due to their complexity and rapid advancement, it has become necessary to accelerate the design and implementation of such systems.

In this context, several code generation works for Multi-Processor Systems-on-Chip (MPSoCs) have appeared in order to facilitate the embedded systems designer task. STARSoC [5] is a framework for hardware/software codesign and design space exploration. The input description consists in a set of software and hardware processes described in C. After specifying the number of processors (instances of the freely available core processor called OpenRisc), a hardware/software partitioning is executed. The hardware part is synthesized in an RTL description and the software part is distributed among the set of processors. As a result, STARSoC generates a bus-based MPSoC platform from a high-level application specification. xENOC [14] is an environment for hardware/software automated design of NoC-based MPSoC architectures. The core of this environment is an EDA tool, called NoCWizard, which can generate RTL Verilog NoCs. The whole system is described in an XML file (NoC features, IPs and mapping), which is used as input for the automatic generation tools. In addition to the hardware infrastructure, xENOC also includes an Embedded Message Passing Interface (eMPI) supporting parallel task communication. NoCMaker [9] derived from xENOC project is an open-source automatic rapid prototyping tool and component re-use environment to build different simulatable, verificable and synthesizable NoC-based MPSoCs in order to perform a design space exploration based on JHDL [10] a Java API for describing and simulate circuits. MAIA [3] is a framework for NoC generation and verification. MAIA can generate different traffic patterns, for different load conditions and source/target pairs. Using the generated traffic and the automatically produced simulation scripts, it is possible to validate and evaluate the NoC and the associated SoC using commercial tools such as the Modelsim simulator. As in xENoC, MAIA can also generate OCP (Open Core Protocol) network interfaces (NIs). NoCGEN [18] is a Network On Chip (NoC) generator, which is used to create a simulatable and synthesizable NoC description. NoCGEN uses a set of modularised router components that can be used to form different routers with a varying number of ports, routing algorithms, data widths and buffer depths. A graph description representing the interconnection between these routers is used to generate a top-level VHDL description. In [6] a rapid prototyping MPSoC based on model-drive approach called LAVA, is presented. The architecture provides a number of open-source IP cores such as PEs (Plasma [15], MB-Lite [16], ZPU [17]), a UART, a timer, and a CAN controller connected to a Wishbone bus [7]. Similar to the xENOC [14], a XML file is used together with VHDL for describing the architecture. However, no synthesizable NoC is provided. SUNMAP [8] is a tool for automatically selecting the best topology for a given application and producing a mapping of cores onto that topology. SUNMAP explores various design objectives such as minimizing average communication delay, area, power dissipation subject to bandwidth and area constraints. The tool supports different routing functions (dimension ordered, minimum-path, traffic splitting) and uses floorplanning information early in the topology selection process to provide feasible mappings. The network components of the chosen NoC are automatically generated using cycle-accurate SystemC soft macros from ×pipesCompiler [12] architecture.

The proposed framework, MppSoCGEN, shares and extends some previous features shown in these related works. It shares the capability of NoCMaker to select the topology, and the capability of NOCGEN to generate top-level VHDL description from user specification of the architecture parameters. Previous systems are not flexible nor scalable to support the requirements of different data parallel applications because they do not support the mixture of different networks. Our proposed framework increases flexibility by using different reconfigurable communication networks based on parametric architecture. MppSoCGEN takes advantage of the automatic generation methods to rapidly design and generate a complete, programmable and flexible SIMD parallel SoC at RTL level dedicated to compute data-intensive applications.

## 3. TARGET MPPSOC ARCHITECTURE

The SIMD FPGA architecture (Fig. 1) is composed of a main controller ACU (Array Controller Unit) connected to its sequential instruction and data memories named InstACU and MemACU respectively, and a number of elementary processors (PEs). Each PE is connected to its local private data memory. We distinguish two mppSoC networks: a neighbourhood interconnection network to assure neighbour communications between PEs and a global router, called a massively parallel Network on Chip, mpNoC, to perform point to point communications. The designer can use none, one or both routers to generate the needed mppSoC configuration. A particularly important feature of the mppSoC is the ability to target diverse applications by customizing the basic architecture. This customization is achieved with its parameterization as well as its extensibility and reconfigurability. In fact, the mppSoC is parametric in terms of the number of PEs as well as the memory size. In addition, the neighbourhood network can be configured at compile time to have one topology among five (2D (mesh, torus, Xnet) and 1D (linear, array)) as illustrated in fig 2.

Furthermore, the mpNoC is designed as an IP which synchronously connects a set of inputs to a set of outputs. It is based on an internal interconnection network that can be chosen at compile time (shared bus, crossbar, multi-stages, etc.) according to the applications requirements. Allowing designer to choose the internal network increases run-time performances.

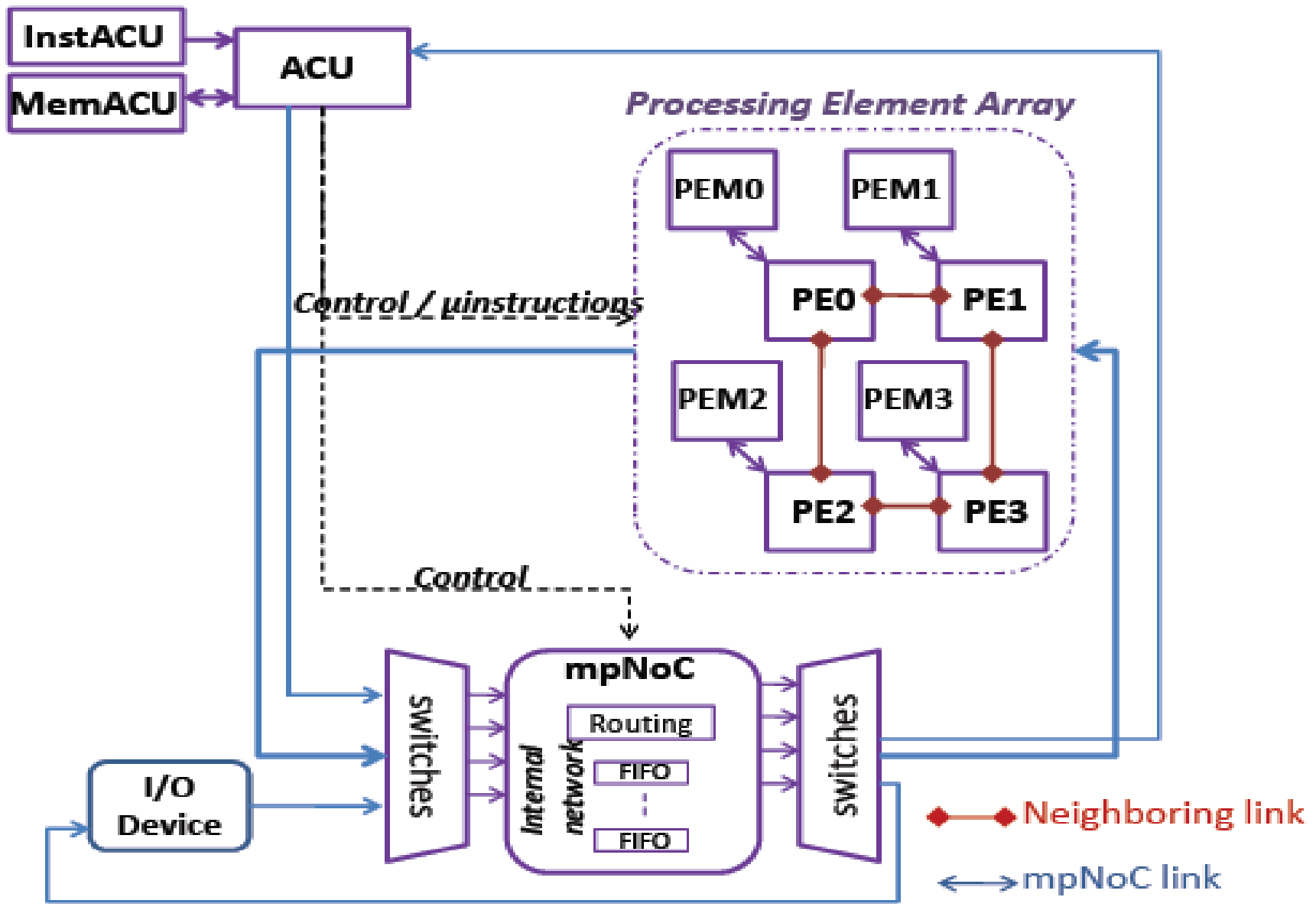

Figure1. Parallel SIMD SoC configuration.

The mpNoC assures irregular communications between processors and also performs parallel I/O data transfer which is clearly a key issue in a SIMD system. The mpNoC can be configured at run-time to support one of the three different bidirectional communication modes (PE-PE, ACUPE, I/O Device-PE) making it powerful and suitable for parallel systems [11].

The architecture is designed as an assembling of various components or IPs, including processors (ACU + PE), memories and networks (neighborhood network and mpNoC). These IPs are partitioned into standard IPs such as processor IP, memory IP and interconnection network IP, and mppSoC dedicated IPs provided by the mppSoC tool. Some standard IPs are furnished in a HW library to alleviate their design.

The designer can also choose his own IP. For that purpose, a descriptive manual can help him to connect one new IP to other mppSoC components. MppSoC dedicated IPs, named IP glues or ad-hoc IPs, must always be used to construct the architecture such as IP controller integrated with the use of a global router to assure the synchronization of the architecture functioning. These IPs are automatically generated by the MppSoCGEN when needed.

To build the mppSoC processors, namely ACU and PE, two assembling methodologies are proposed. The first one, called reduction methodology, is based on the reduction of the main processor in order to have a small reduced one. This significant gain allows integrating a large number of PE on a single chip. In the second methodology, called replication methodology, the PE is constructed by the same processor as the ACU to reduce the design time and make the architecture assembling faster. This methodology offers a large gain in the development time since it is easy. However, we are unable in this case to integrate many PEs on a single chip. The criterion for using this methodology is to choose a smaller processor that can be fitted in large quantities in one FPGA. So, there is a compromise between the two processor assembling methodologies and the designer has to choose the most appropriate one satisfying his needs.

In our generated mppSoC configuration, they are three main VHDL files:

- « user_library »: include the declaration of different used entities in the implementation as well as the definition of the new types and functions adapted to mppSoC.

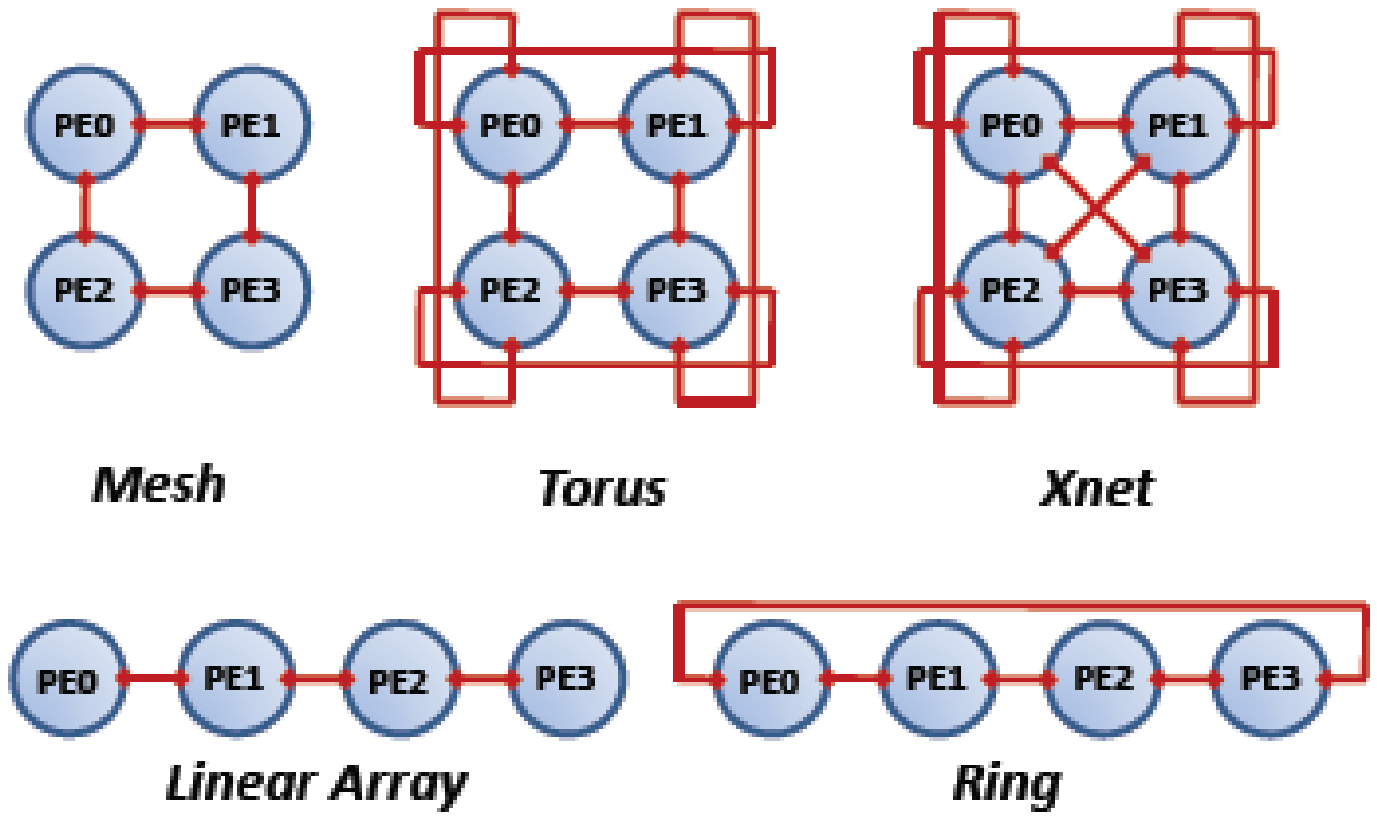

Figure2. Neighbourhood network topologies

- « pack_mppsoc» : present the configuration file of mppSoC. It include the declaration of different constants (the arrangement and number of PEs (number of rows, number of columns (sl_nb_rows , sl_nb_column) , the memory size (ACU/PE) address memory (MS_add_width, SL_add_width) , the topology of the neighborhood network if exists (linear, ring, mesh, torus, xnet)). Moreover, we find in this file a declaration of the components appropriate for the used processor.
- « mapping_mppsoc » : the principal file that define the mppSoC architecture (top level).

In order to generate one mppSoC configuration, MppSoCGEN has been developed. It helps the designer choosing a SIMD configuration at a high abstraction level and automatically generating its source code. The generated system can be then directly prototyped or simulated using synthesis and simulation tools (such as the Quartus synthesis tool for Altera devices and Modelsim tool for simulation). The next section presents the MppSoCGEN design flow.

## 4. MPPSOCGEN DESIGN FLOW

MppSoCGEN offers a flexible design flow to automatically generate an mppSoC architecture model at an RTL abstraction level as specified by the designer at a high abstraction level. An outline of the framework's functionalities is shown in fig4. MppSoCGEN design flow addresses two key problems in mppSoC configuration: the need for tools to quickly and efficiently generate mppSoC architecture, and the requirement for configuration validation to assure the generated architecture performance.

The starting point of the design flow is the user specification of the application's requirements to create an mppSoC configuration. The framework uses an IP library with various components (processors, memories, interconnection networks...) that can be selected in the deployment process to generate the needed SIMD configuration.

To assure the efficiency of the mppSoC configuration, a verification of configuration step is needed. It consist of executing some basic rules, which assures, that the design specifications of the user make sense, e.g. a 2D mesh communication infrastructure is not allowed if only one row of PEs was chosen. Once the entire mppSoC configuration is ready, the framework modifies VHDL libraries based on parsing method to set the appropriate VHDL parameters and to generate the desired configuration. A synthesizable RTL VHDL description of the MppSoC is then produced. Finally, as denoted in Fig.4 below the dashed line, commercial software tools, i.e., ModelSim Altera, and Quartus II for the design simulation and synthesis onto FPGA devices are used in order to provide the user information about the design implementation; namely, the design size in terms of resource utilization, the time execution, etc.

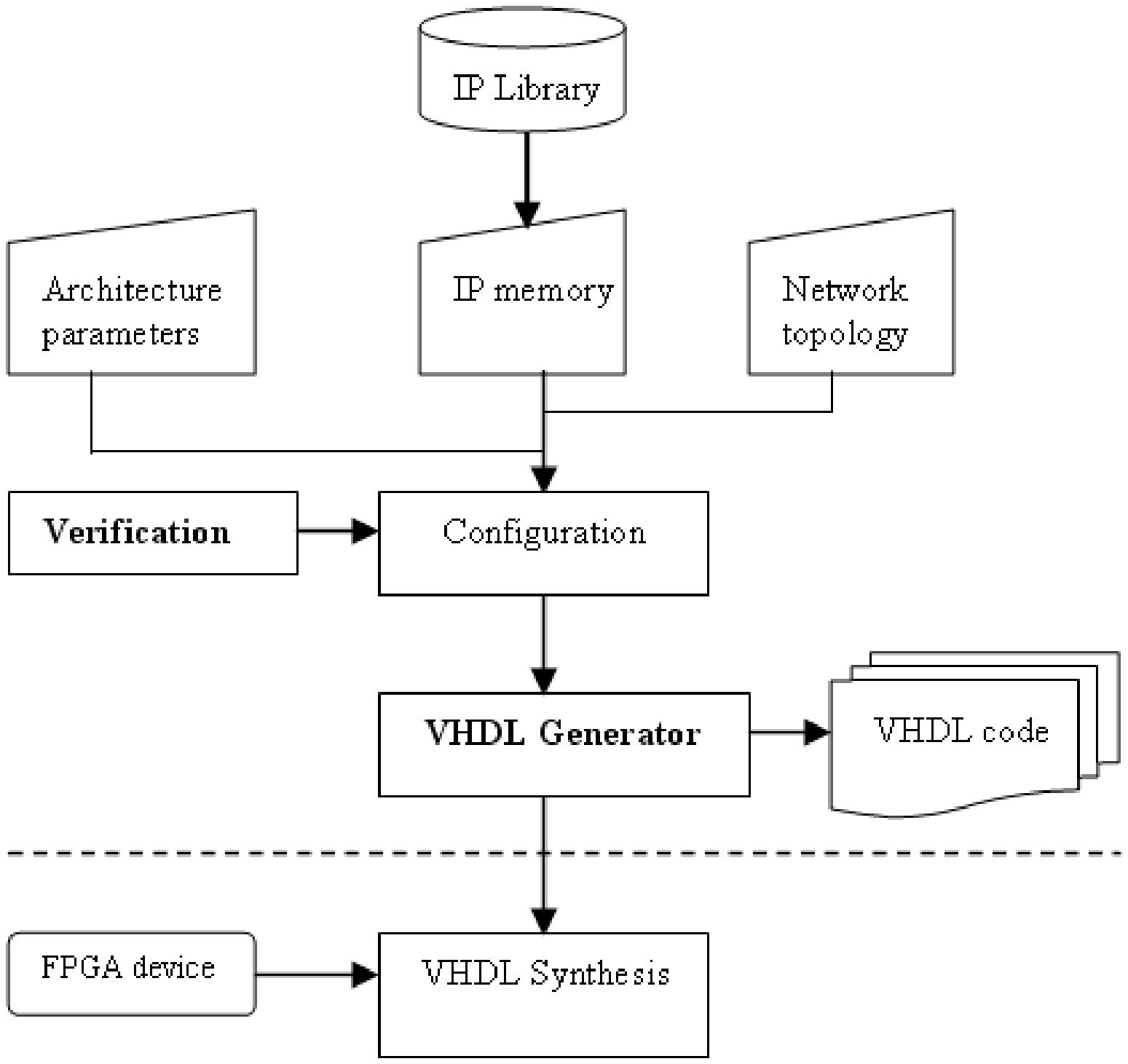

Figure 3. MppSoCGEN design flow

The designer can generate a SIMD massively parallel SoC configuration in three steps: user specification, mppSoC configuration verification, and code generation.

### *4.1. User specification*

The user specification gathers a set of concepts to specify the application part of a system. In our design flow, user can create his proper mppSoC configuration at high abstraction levels through a simple Graphical User Interface (GUI) to generate its VHDL code. Since mppSoC is designed as a parametric architecture, the designer has to set some parameters in order to generate one configuration on FPGA (fig.4):

1. Select the IP processor from the mppSoC IP library (we can actually choose among three available processors: minimips[21], MIPS[22] and NIOS[23]);
2. Choose the processor assembling methodology (reduction/replication);
3. Choose the mppSoC architectural parameters namely the number of PEs and their arrangement (1D or 2D grid) as well as the memory size;
4. Select/no the neighborhood network and its topology. In this step the user can choose among linear or ring topologies if he has chosen a 1D PE arrangement or among torus, mesh or Xnet if the PEs are placed in a 2D grid;
5. Select/no the mpNoC interconnection network (shared bus, crossbar or Delta multi-stage network (omega, baseline or butterfly))
6. Choose the IP memory from the IP library;

A help manual is in fact provided to the designer to describe the different instructions to use according to the chosen processor.

By this way, the designer can generate the configuration satisfying his requirements. We notice that our framework facilitates the mppSoC design and that any modification in the configuration could be easily done.

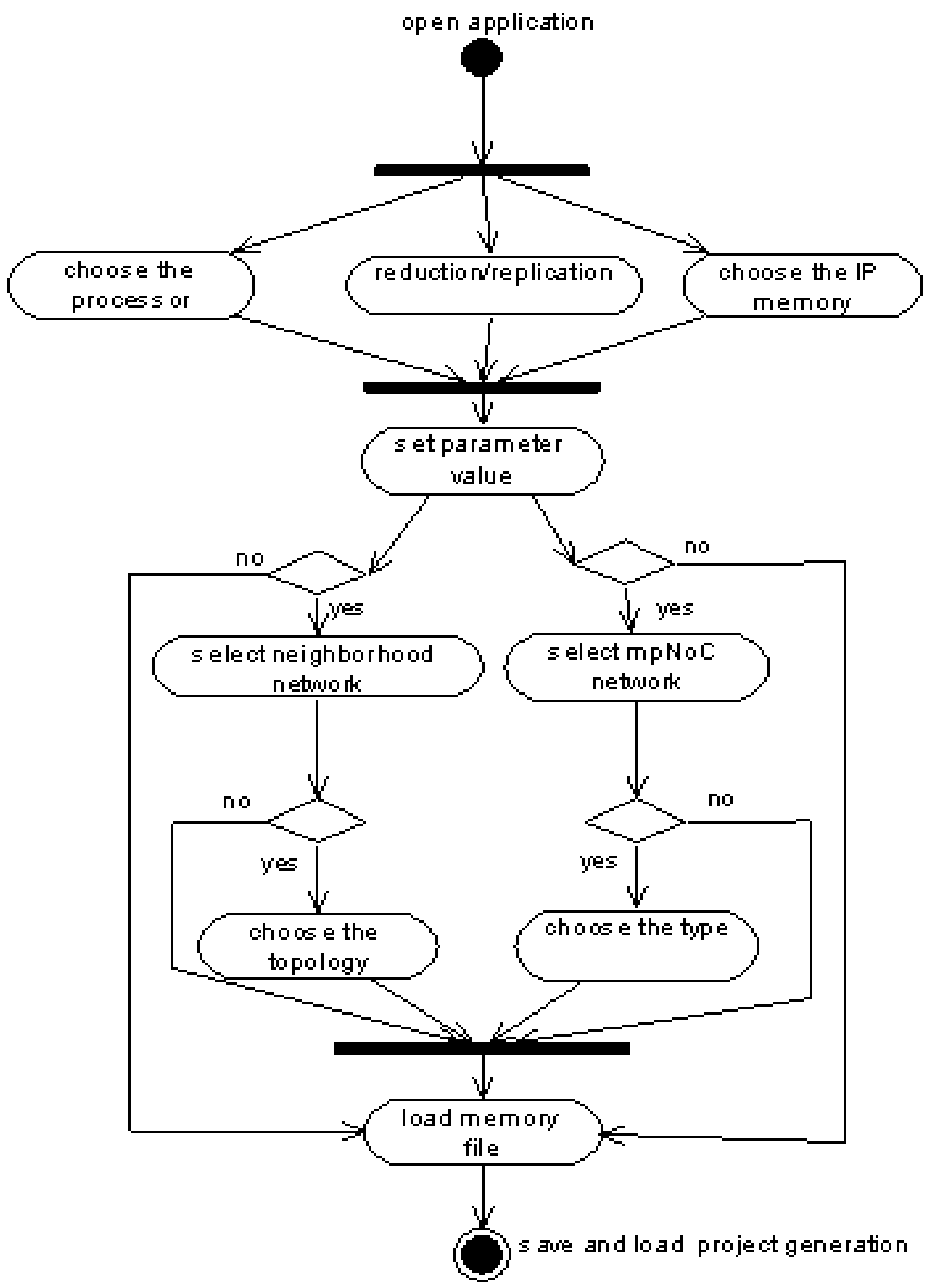

Figure 4. UML Activity diagram to create an mppSoC configuration

### *Verification of MppSoC configuration*

As a guaranty of the mppSoC configuration high quality, a verification process is applied. It checks if the proposed configuration meets a set of mppSoC design specifications and regulations by applying some configuration rules. Table 1 illustrates some rules for verifying the mppSoC configuration.

This section describes the verification of mppSoC configuration algorithm (Algorithm 1) which executes described configuration rules. Algorithm 1 defines essentially integer variables sl_nb_rows , sl_nb_column which describe the PEs arrangement, String variables mpNoC_type, neighboring_type which correspond to the mpNoC and neighborhood network topologies and a boolean variable isvalid that describe the validation of the mppSoC configuration. The configuration become invalid, when isvalid takes false.

Despite its simplicity, Algorithm1 illustrates two principal advantages. First, it has a very low complexity compared to other verification algorithm [28]. Second, it quickly achieves good code coverage: in 10 µs it executes 10 tests.

Table 1: mppSoC configuration rules

| Rules | Description |
|---|---|
| R1 | if (PE number rows and PE number columns $\neq 2^n$), then the Delta MIN (Multistage Interconnection Network) mpNoC interconnection network type should not be chosen |
| R2 | if (PE number rows=1), then only Linear and Ring neighboring network topologies are valid. |
| R3 | if (PE number row $\succ$ 1), then only mesh 2D, Torus 2D and Xnet neighboring network topologies are valid. |

### *Code generation*

A parsing method is proposed to automatically generate a SIMD configuration at a high abstraction level. Indeed, to generate the VHDL files with the new parameters values, we need to analyze and modify VHDL code. Parsing is the division of text into a set of discrete parts, or tokens, which in a certain sequence can convey a semantic meaning. It is then used to extract structural information from the source code.

To assure the efficiency of software code generation, some parsing rules are implemented. Executing these rules, a correct and suitable VHDL description of the mppSoC configuration is generated. Parsing problems are classified as follows:

***Problem1: Generate the pack_mppsoc.vhd file with the new parameters values introduced in the GUI.***

- Rule1: The VHDL line must start with "constant" expression.
- Rule2: the parameter value must be written immediately next the ":=" token.

***Problem2: Generate the pack_mppsoc.vhd file with the chosen* neighboring *topology in the GUI.***

- Rule1: The VHDL line must start with "constant" expression.
- Rule2: the parameter value must be written immediately next the ":=" token.

***Problem3: Generate the VHDL memory file by inserting the name of the selected memory file in the GUI.***

- Rule1: The VHDL line must start with "init_file" expression.
- Rule2: the selected memory file name must be written immediately next the "=>" token.

***Problem4: Generate the VHDL memory file with the new parameters values introduced in the GUI.***

- Rule1: The VHDL line must start with "address" expression and the parameter value must be written immediately next the "STD_LOGIC_VECTOR" expression.
- Rule2: The VHDL line must start with "numwords_a" expression and the parameter value must be written immediately next the "=>" token.
- Rule3: The VHDL line must start with "widthad_a" expression and the parameter value must be written immediately next the "=>" token.

**Algorithm 1: verification of Configuration algorithm**

```
iSvalid <- true;
[Rule1: R1]
if sl_nb_rows * sl_nb_columns = 2^n
    if mpNoC_type= "Delta Min" then
    iSvalid <-false; Break;
    end if;
end if;
[Rule2: R2]
if sl_nb_rows=1
    if neighboring_ type= "mesh 2D"
    iSvalid <-false; Break;
    end if
        else if neighboring_ type= "Torus 2D"
    iSvalid <-false; Break;
    end else if
    else if neighboring_ type= "Xnet"
    iSvalid <-false; Break;
    end else if
end if
[Rule3:R3]
else if sl_nb_rows >1
    if neighboring_ type= "Linear"
    iSvalid <-false; Break;
    end if
    else if neighboring_ type= "Ring"
    iSvalid <-false; Break;
    end else if
end else if
else
    iSvalid <-false; Break;
end else
```

**Agorithm2: Lexer1 java code**

**Step 1: [Reading VHDL Code]**

```
BufferedReader in=new BufferedReader(VHDLfile);
while((line=in.readLine())!=null){
numline=numline+1;
```

**Step 2: [Selecting line]**

```
if (line.contains("constant")) {
```

**Step 3:[generation of tokens]**

```
StringTokenizer st = new StringTokenizer(line);
```

**Step 4: [Tokens parsing]**

```
while (st.hasMoreTokens()) {
tmp=st.nextToken();
if(tmp.equalsIgnoreCase(VariableName)){
while(!st.nextToken().equals(":=")&&
st.hasMoreTokens())
{ }
if (st.hasMoreTokens()) { value=st.nextToken();
```

**Step 5: [line update]**

```
line=line.replaceFirst(value, newVarValue);
list_numlines.add(numline);
list_lines.add(line); } } } }
```

**Step 6: Finish**.

In the developed code generation program, a specific parsing algorithm is implemented using StringTokenizer java class for each previously described problem. All algorithms have the same principle implementation but they differ in how insert the new variables values. Algorithm 2 describes how MppSoCGEN works generating pack_mppsoc.vhd file with the new parameters values chosen by the user (problem1). The entire code generation process consists of three phases: 1) Reading the VHDL file and selecting VHDL code lines for update. 2) Generating and parsing tokens. 3) Tokens updating and generating new VHDL code. Fig.2 shows the proposed parsing method to generate a complete VHDL code.

To use StringTokenizer, an input string and a string that contains delimiters must be specified. Delimiters are characters that separate tokens. Each character in the delimiters string is considered a valid delimiter, for example, ",;:" sets the delimiters to a comma, semicolon, and colon. The default set of delimiters consists of the whitespace characters: space, tab, newline, and carriage return. Once a StringTokenizer object is created, the nextToken( ) method is used to extract consecutive tokens. The hasMoreTokens( ) method returns true while there are more tokens to be extracted. Algorithm2 analyzes separately each code line to search which line must be updated. After an analysis test, VHDL code line is selected for updating. The StringTokenizer class accepts the string of input line to demarcate it into sequence of tokens that can be presented. Each token is then analyzed to know the position of the new parameter value. After updating, the input line will be inserted into the original VHDL code. Thus, the process continues until we obtain the final VHDL code.

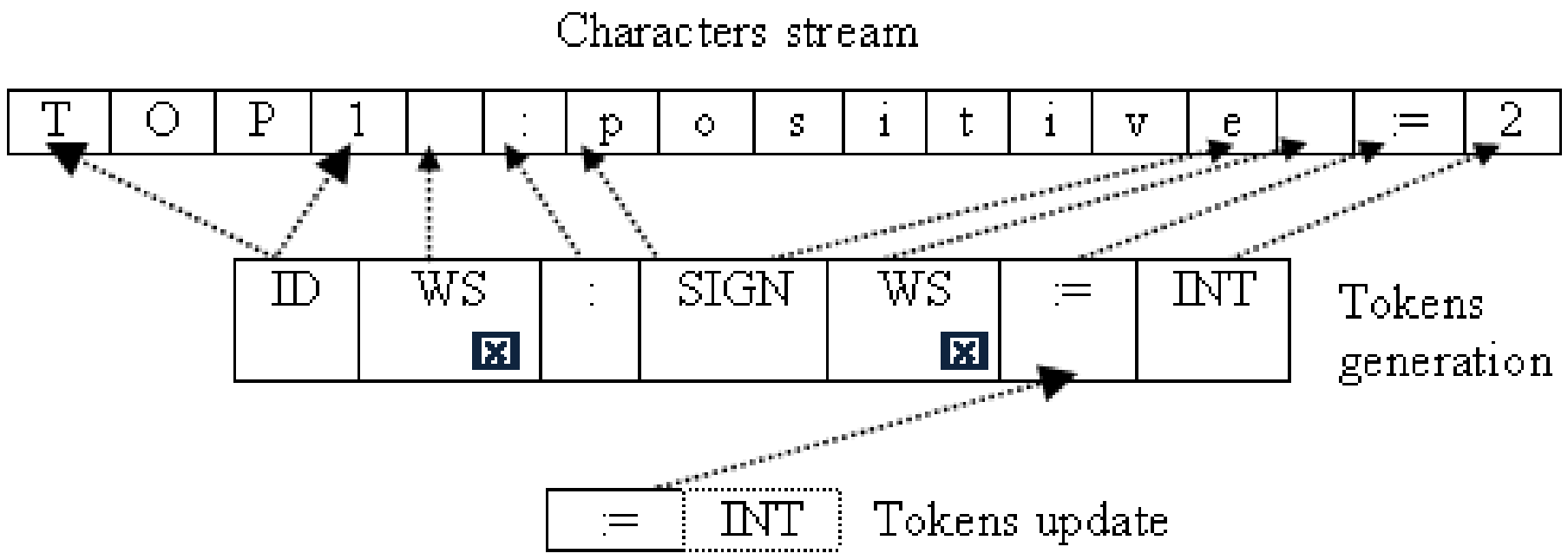

Figure 5. Parsing method

Algorithm2 illustrates two original features. First, it has small size – 15 lines (low complexity). Second, it quickly achieves good code coverage: in 1 ms it automatically update and generate the pack_mppsoc.vhd file.

Executing its low_comlexity algorithms, MppSoCGEN is capable of rapidly generating 2566 lines of VHDL code with only 300 lines of source code. If we compare the numbers presented in table 2 with other java generator tool, the results are still very good. Indeed, our framework improves its flexibility on Netbeans 5.5 version and centrino duo Core 2GHz with 22 Kbytes and 3 seconds average runtime while GenERTiCA [29] works with 33 Kbytes and 10 seconds average runtime. The next section illustrates the use of MppSoCGEN in a real application context.

Table 2: Complexity comparison

| | **MppSoCGEN** | **GenERTiCA [29]** |
|---|---|---|
| **Lines of source code** | 300 | 388 |
| **Used memory** | 21 KB | 33 KB |
| **Runtime** | 3s | 10s |
| **Generated files** | 38 | 21 |
| **Generated lines of code** | 2566 | 1264 |

## 5.CASE STUDY

The MppSoCGEN design flow presented above has been used to generate many mppSoC architectures. We present here a case study on reduction algorithm [19] that demonstrates the efficiency of the proposed design flow. The design was synthesized for the FPGA Stratix 2S180 device which includes 143520 ALUTs for hardware logic [24]. Experimental results were obtained using the ModelSim Altera simulator to simulate and debug the automatically generated design and the Quartus II which is a synthesis and implementation tool [31] used also to download the compiled program file onto the chip.

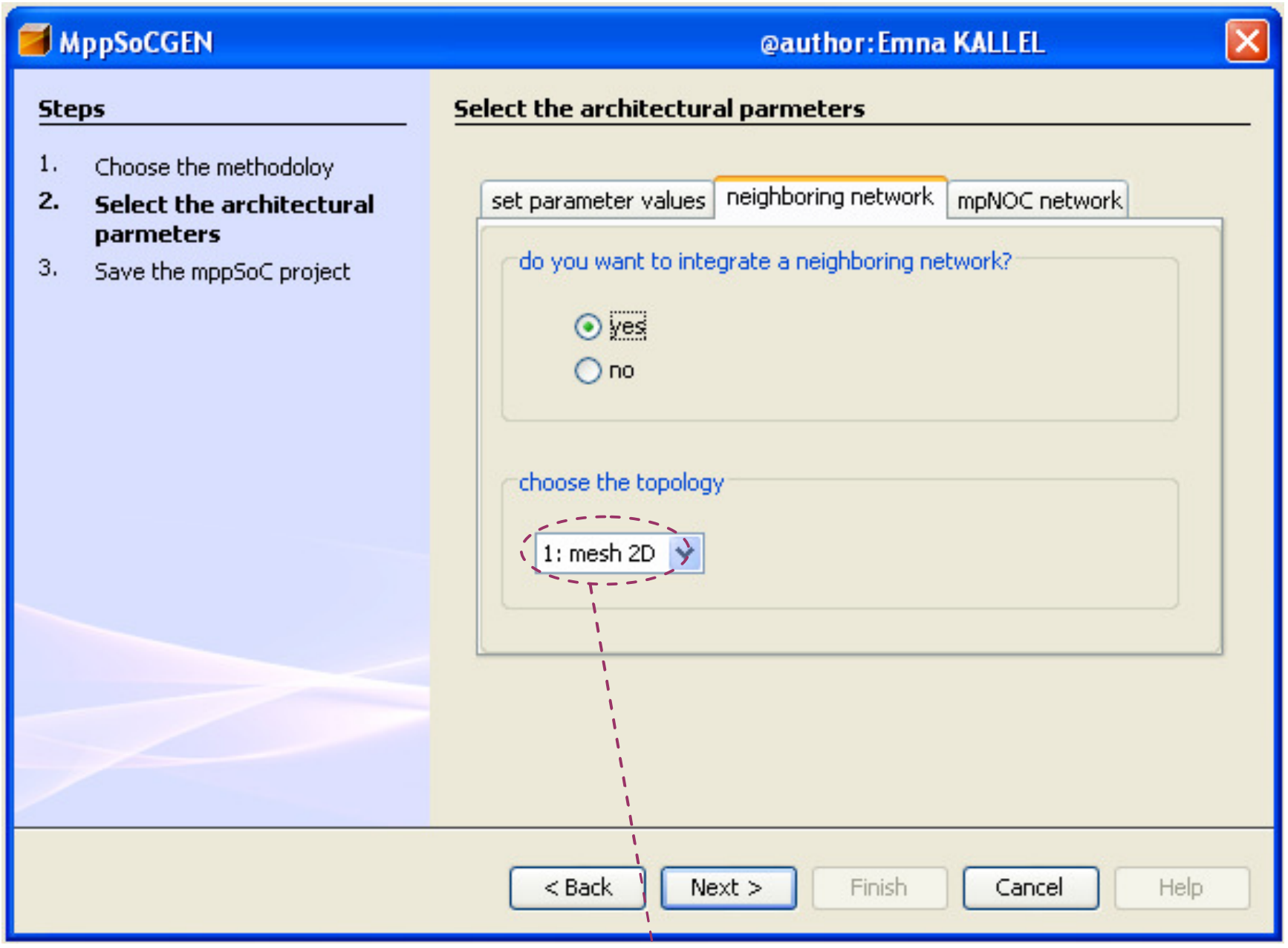

```
--Reg Router Topologies
-- 0: no router, 1: mesh 2D, 2: Torus 2D, 3: Xnet,
-- 4: Linear, 5: array (Ring)
constant Top1         : positive := 1 ;   -- Mesh topology
```

Figure 6. Code generation example

The reduction algorithm presents one basic image processing operations. When reduction computation is conducted in parallel, it is known that the computation can be completed with the minimum number of steps using a binary tree representation. To implement the reduction algorithm we use the recursive doubling procedure, sometimes also called tree summing. This algorithm combines a set of operands distributed across PEs [20]. Consider the example of finding the sum of M numbers. Sequentially, this requires one load and M-1 additions, or approximately M additions. However, if these M numbers are distributed across N = M PEs, the parallel summing procedure requires log2(N) transfer-add steps, where a transfer-add is composed of the transfer of a partial sum to the PE and the addition of that partial sum to the PE's local sum. The generated configurations could be directly simulated to measure execution time and decide the performance of the SIMD modeled systems. Fig.6 shows VHDL code generation example specifying a 2D mesh neighboring network from MppSoCGEN GUI.

Execution performances are compared (Fig.7). We note that the architecture of the used parallel system has also a great impact on the speedup of a given algorithm. We notice that the mppSoC based on the regular network is the most effective for this type of application. In the case of a completely-connected topology, the speedup is six times lower than when using a mesh inter-PE network. The two regular topologies, mesh as well as linear array, give approximately the same execution time. Indeed, the time obtained with a linear router is slightly lower than with a mesh router. This is due to the additional communication overhead introduced by the mesh router. So, the linear neighborhood network is the most effective of the reduction algorithm. These different results show also the flexibility of the mppSoC architecture and the high efficiency achieved by establishing a well mapped network topology to one algorithm.

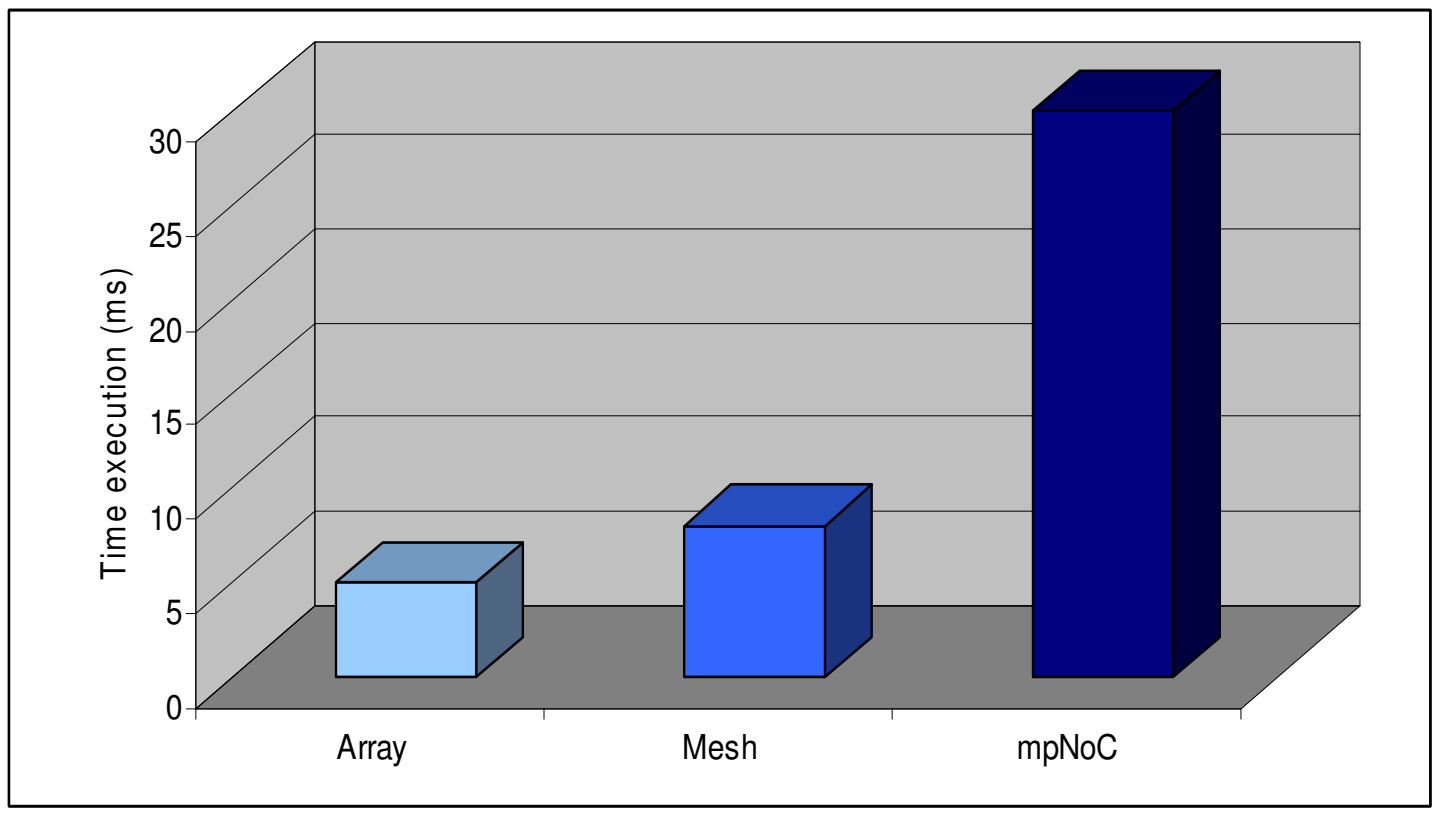

Figure 7. Execution time on different mppSoC configurations.

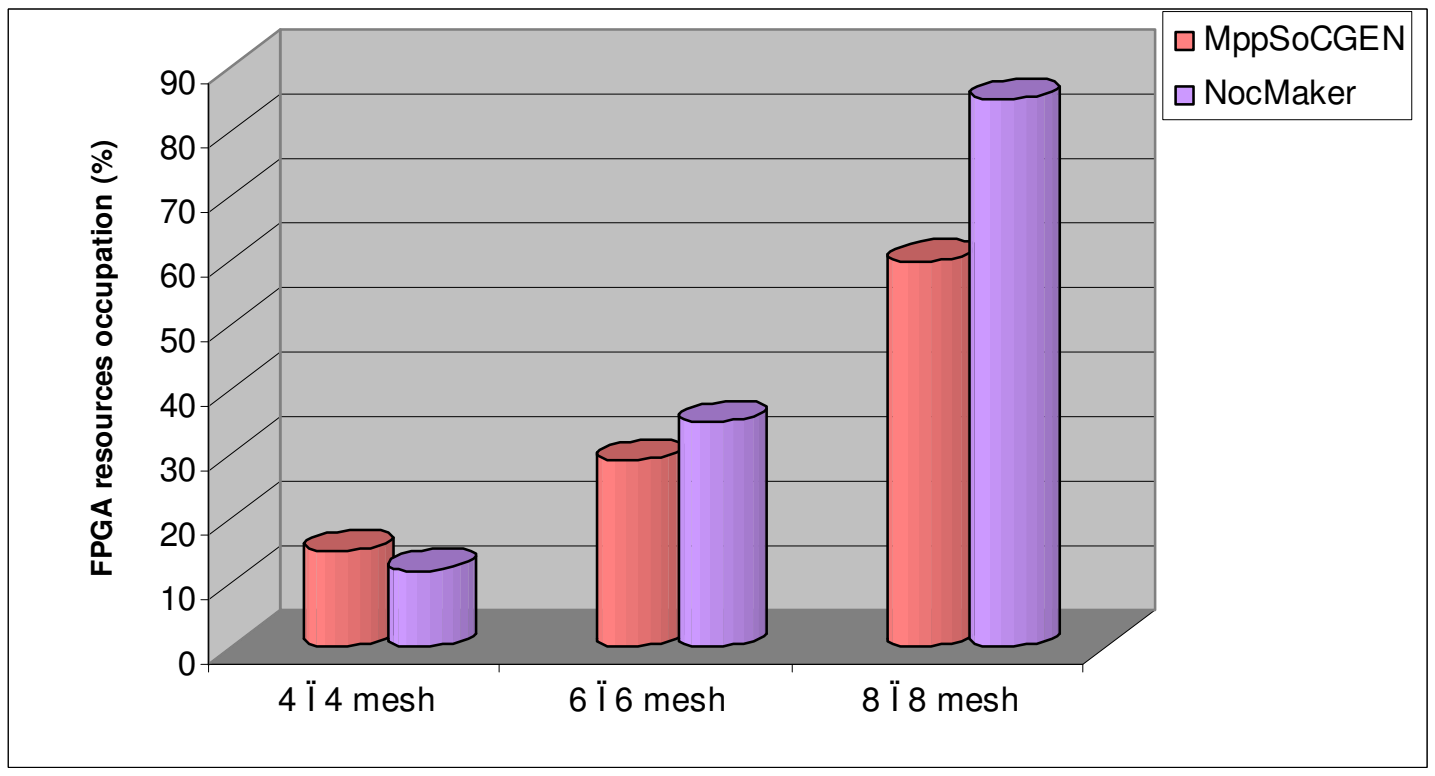

Figure 8. Performance comparison running a reduction algorithm on a Stratix 2S180 (2D mesh)

Table 3 shows the obtained synthesis results varying the SIMD parameters. We clearly notice that the processor replication consumes more FPGA area than the processor reduction. With the latter methodology, we can put a large number of PEs on a single chip (up to 84 PEs on the Stratix 2S180). We also deduce the impact of the integrated network in the FPGA resources. Comparing for example between the first and the second configurations, it is clearly shown that the crossbar consumes more FPGA logic than the Delta multi-stage network. The neighbourhood network also consumes fewer FPGA resources than the mpNoC. Thus, depending on his needs the designer can integrate the needed components in the selected mppSoC configuration. The mppSoC generation tool significantly facilitates the mppSoC design and rapidly allows the modification of the SIMD configuration. It just takes fewer seconds to generate an mppSoC configuration.

In order to prove the performance of MppSoCGEN, Fig.6 presents the obtained FPGA resources allocation comparative results using MppSoCGEN and NoCMaker [9][14] frameworks. The topology used in this comparison is 2D mesh since NocMaker only support the mesh topology. We notice comparable results between the two frameworks on 4×4 mesh topology. However, when increasing the number of PEs, the FPGA area is multiplied by a factor of 4 using MppSoCGEN which is an acceptable rate while the FPGA area using NoCMaker is multiplied by a factor of 7 which is a high rate compared to the MppSoCGEN results. This comparison proves the high performance and the scalability of MppSoCGEN framework.

Table 3: Synthesis results

| PEs | Processor Assembling methodology | Neighbourhood Network | MpNoC Inter-connection Network | Logic utilization | | | Total memory | | |
|---|---|---|---|---|---|---|---|---|---|
| | | | | *ALUTs* | *registers* | *%* | *ACU (bytes)* | *PE (bytes)* | *%* |
| 8 | Replication | _ | Crossbar | 42059 | 16863 | 27 | 4096 | 1024 | 3 |
| 16 | Reduction | _ | Crossbar | 27268 | 11481 | 21 | 4096 | 4096 | 6 |
| 32 | Reduction | 2D Torus | Crossbar | 81796 | 22478 | 45 | 2048 | 600 | 4 |
| 64 | Reduction | Xnet | _ | 89154 | 24479 | 49 | 4096 | 2048 | 8 |
| 84 | Reduction | _ | Delta Min | 130603 | 41620 | 96 | 2048 | 2048 | 10 |

Table 4: difference between Manuel and automatic solutions

| methodology | Configuration Time (automatic) | Configuration Time (manual) |
|---|---|---|
| Reduction methodology | 12 minutes | 744 hours |
| Replication methodology | 40 seconds | 168 hours |

Also, in order to illustrate the efficiency of our framework, Table 4 compares the implementation design time using MppSoCGEN with results obtained from a conventional manual implementation method done by the same designer without using any framework. The measured design time for the second configuration (using replication methodology) is just the time needed to modify the first configuration (with reduction methodology). The results in Table 4 show that the proposed framework is a better solution to accelerate the design of specific SIMD parallel SoC according to the estimated design time compared to a manual design. One month was necessary to reduce an open-source processor to obtain a small PE (with only execution units) [30]. Observing the results, we can conclude that the MppSoCGEN allows a very fast SIMD implementation.

From all previous experiments we demonstrate the effectiveness and the flexibility of the generated mppSoC systems compared to previous similar systems and the capacity of MppSoCGEN to facilitate and accelerate SIMD SoC implementation to run data-parallel applications. Indeed, through our MppSoCGEN design flow presented in this work, a designer can select the needed SIMD configuration at high abstraction level and automatically generate its implementation at RTL high abstraction level. The designer can easily and rapidly generate different SoC configurations to look for the best alternative for a given application.

## 6. CONCLUSION

An mppSoC generator tool for SIMD SoC configuration, MppSoCGEN, was presented. The proposed design flow is composed of three steps: user specification, verification of parallel soc configuration, code generation. Using our flexible MppSoCGEN framework, the SIMD SoC design is accelerated. The VHDL implementation can be automatically generated based on parsing method. MppSoCGEN also facilitates SoC exploration and helps the user choose the best configuration for a given application. The framework flexibility was improved on Netbeans 5.5 version and centrino duo Core 2GHz with 22 Kbytes and 3 seconds average runtime. Experimental results show that the proposed framework strongly contributes to the increase of the designer's productivity. The case study with signal processing application proved that the presented design flow can facilitate and accelerate the design of parallel SIMD SoC systems.

The adopted mppSoC architecture allows reducing implementation costs. Moreover, our parsing method makes the mppSoC configuration easier and faster than a manual configuration.

As future work, the developed framework will be enhanced to support automatic exploration. A high-level exploration mppSoC exploration step could help the user generate the best configuration for a given data-parallel application.

## ACKNOWLEDGEMENTS

We would like to thank Dr. Raouf Ketata for his helpful review of the manuscript.

## Authors

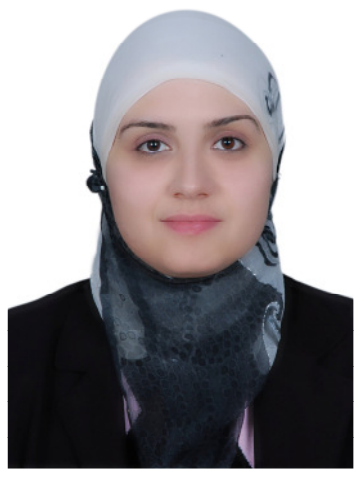

Emna KALLEL: received his Dipl.-Ing. degree in 2006 from the National School of Engineers of Sfax (ENIS) Tunisia, where he is currently working toward the Ph.D.degree with research focused on automatic code generation for embedded systems. She has been with the higher institution of electronics and communications of Sfax, Tunisia, as Research Assistant.
Here further research interests include parallel computing, parallel architectures, automatic parallelisation, Rapid prototyping and video processing and object-oriented methods for hardware generation.

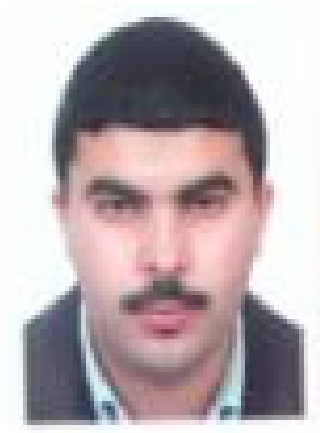

Yassine Aoudni: Studied Electrical Engineering and Computer software Engineering at the National School of Engineers of Sfax (ENIS) Tunisia. He received Dipl.-Ing.from the National School of Engineers of Sfax in 2002 and Dr.-Ing. from the University of South Brittany**,** France in 2010. From 2008 till 2011, he worked as a Research Assistant at National School of Engineers of Sfax (ENIS) Tunisia conducting research in FPGA prototyping Since 2011 he is Assistant Professor at the National School of Engineers of Sfax.
He acts as a member of several technical program committees, as a reviewer for different journals. His research interests include joint source FPGA prototyping, signal processing, system high level design, parser design, information theory, as well as multiprocessor architecture.

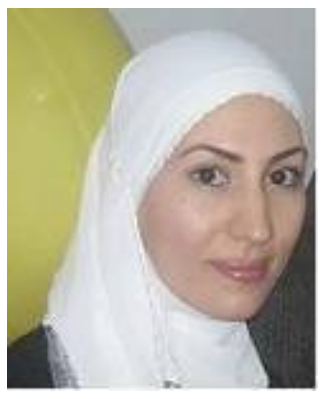

Mouna Baklouti: She received Dipl.-Ing. From the Tunisia polytechnic school in 2005 and Dr.-Ing. from the National School of Engineers of Sfax (ENIS)**,** Tunisia in 2010. From 2007 till 2011, she worked as a Research Assistant at National School of Engineers of Sfax (ENIS) Tunisia conducting research in parallel computing. Since 2011 she is Assistant Professor at the National School of Engineers of Sfax.
She acts as a member of several technical program committees, as a reviewer for different journals. Here research interests include joint source parallel architecture, signal processing, system high level design, Interconnection network, as well as video coding.

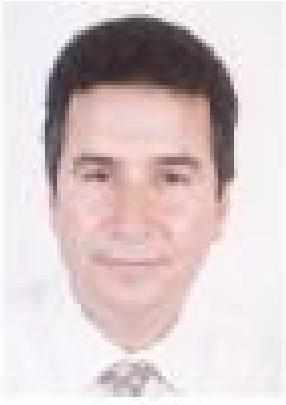

Mohamed Abid: He received Dipl.-Ing. from the National School of Engineers of Sfax (ENIS) in 1985 and the phd degree from the National Institution of applied Science, Toulouse, France. In 2000, he received his doctorale degree in Electrical and Computer Engineering at National Engineering School of Tunis. He is currently Professor at the Electrical Department of ENIS.
Since 2006, he has been on the Head of the research laboratory «Computer Embedded System » CES-ENIS. He is responsible for research projects in the area of automatic signal and image processing, wireless networks and information systems. He has been on the Head of Federator Research Project since 2009. He has authored or co-authored more than 120 international conference papers, and he has written more than 20 technical contributions to various international standardization projects. He is a member of the Scientific and Program Committees of several international conferences and workshops. He is the Co-coordinator of several Nationals and Internationals projects with universities and industries like DGRSRT, CNRS, INRIA, CMCU, training for research, PNM, Tempra, etc.

.